# Non-altermagnetic spin texture in MnTe

Meng Zeng[1,*], Pengfei Liu[1,*], Ming-Yuan Zhu[1,*], Naifu Zheng[1], Xiang-Rui Liu[1], Yu-Peng Zhu[1], Tian-Hao Shao[1], Yu-Jie Hao[1], Xiao-Ming Ma[1], Gexing Qu[2], Rafał Kurleto[3], Dawid Wutke[3], Rong-Hao Luo[1], Yue Dai[1], Xiaoqian Zhang[4], Koji Miyamoto[5], Kenya Shimada[5], Taichi Okuda[5], Kiyohisa Tanaka,[6] Yaobo Huang,[7] Qihang Liu[1,†], Chang Liu[1,†]

[1]*Department of Physics, State Key Laboratory of Quantum Functional Materials, and Guangdong Basic Research Center of Excellence for Quantum Science, Southern University of Science and Technology (SUSTech), Shenzhen, Guangdong 518055, China*

[2]*Beijing National Laboratory for Condensed Matter Physics and Institute of Physics, Chinese Academy of Sciences, Beijing 100190, China*

[3]*SOLARIS National Synchrotron Radiation Centre, ul. Czerwone Maki 98, 30-392 Kraków, Poland*

[4]*Key Laboratory of Quantum Materials and Devices of Ministry of Education, School of Physics, Southeast University, Nanjing, Jiangsu 211189, China*

[5]*Hiroshima Synchrotron Radiation Centre, Hiroshima University, Higashi-Hiroshima, Hiroshima 739-0046, Japan*

[6]*Institute for Molecular Science, National Institutes of Natural Sciences, Myodaiji, Okazaki 444-8585, Japan*

[7]*Shanghai Synchrotron Radiation Facility, Shanghai Advanced Research Institute, Chinese Academy of Sciences, Shanghai 201204, China*

# Abstract

Recently, altermagnets have emerged as promising candidates in spintronics, uniquely combining large spin-polarized electronic states with zero net magnetization[1–7]. A prominent example is $\alpha$-MnTe, whose altermagnetic spin splitting, i.e., the degeneracy lift in momentum space induced by collinear magnetic order, has been experimentally observed[8–12]. However, the direct evidence of its *g*-wave spin polarization, the key property for altermagnetic spintronics, is thus far lacking. By combining high-resolution spin- and angle-resolved photoemission spectroscopy (SARPES) with first-principles calculations, we reveal a $k_z$-independent, Rashba-like spin texture in $\alpha$-MnTe. Our results indicate that the observed spin polarization is primarily governed by spin-orbit coupling, whereas the magnetic order contributes to the splitting of energy bands but plays a much less dominant role in spin polarization due to the multi-domain nature[13]. From this result, we further establish a way to prescreen altermagnet candidates that favor the formation of large antiferromagnetic domains based on symmetry analysis. Our work elucidates the interplay between magnetic order and spin-orbit coupling in governing spin polarization in altermagnet candidates, and thereby advances the materials design paradigm for spin-functional devices.

Modern spintronics lies in the utilization of spin polarized electronic states for the manipulation of spin degrees of freedom in solid-state systems[14–18]. Indeed, the concept of spin polarized electronic state consists of two essential physical quantities: spin splitting, splitting of the otherwise degenerate energy bands due to the symmetry breaking by magnetic order or spin-orbit coupling (SOC), and spin polarization, electronic spin carried by the Bloch wave functions of the split bands[19]. Obviously, both ingredients are indispensable for realizing spintronics applications. In ideal models, spin splitting often corresponds to complete spin polarization with characteristic spin-momentum locking patterns. For example, the SOC-induced Rashba spin splitting results in in-plane helical spin textures[20,21]. On the other hand, altermagnets manifest momentum-dependent spin splitting induced by the collinear anti-parallel local moments distributed within anisotropic crystalline fields[1–7]. In the nonrelativistic limit without SOC, the azimuth spin angular momentum serves as a good quantum number, rendering fully spin-polarized states for the spin-split bands.

However, spin polarization may not be robust in real materials. For example, SOC can mix different orbital components in the spin-split bands, yielding a truncated spin polarization[22–25]. In magnetic materials, the presence of magnetic domains, often connected by symmetry, tends to average the spin signals of independent domains[26]. This averaging significantly reduces the level of spin polarization in the spin split bands, hampering the production of spin currents and/or Hall voltages in realistic altermagnet-based spintronic devices[27]. On the other hand, the SOC effect, while often considered a secondary effect in altermagnets, is immune to multidomain structures linked by rotational symmetry. Therefore, experimental observation of not only the spin splitting but also the spin polarization, and elucidation of the origin of spin polarization, is crucial for the application of altermagnets.

MnTe is a prototypical altermagnet whose pronounced spin-split bands and multidomain structure are observed experimentally[13], rendering an ideal platform to examine its magnitude of spin polarization and the underlying mechanism. A number of spectroscopic experiments reveal the lift of spin degeneracy of bulk bands[8–12]. One of these works provided additional information on the spin of these bands[10]. However, comprehensive knowledge of its 3D configuration of spin polarization, and thus direct evidence of its proposed *g*-wave altermagnetic spin texture, is by far lacking due to insufficient momentum points selected for spin-resolved measurements in the 3D BZ. In this study, we examine the spin polarization texture of MnTe by employing systematic SARPES measurements and DFT calculations. In contrast to the *g*-wave spin texture within the altermagnetism scenario, our data reveals that the sole reliable spin polarization signal in MnTe originates from the contribution of SOC in the surface states. The MnTe bands near the Fermi level ($E_F$) exhibit a spin-momentum-locked behavior, where the chirality of the in-plane spin remains unchanged for a specific band regardless of both $k_\parallel$ and $k_z$, similar to those found in the surface states of topological insulators[28,29]. On the other hand, no conclusive spin polarization is detected in its bulk bands allocated further away from $E_F$. Our DFT calculation suggests that such spin configuration results from the SOC-induced spin polarization on the surface electronic states, which is left observable when differently-oriented antiferromagnetic domains are simultaneously probed in SARPES experiments. Our study unveils a unique reduction of altermagnetic spin polarization despite the presence of altermagnetic band lifting, and demonstrates that in realistic, multidomain altermagnetic materials, SOC could potentially dominate the spin-related transport and greatly affect the realization of the proposed exotic quantum phenomena. In addition, we show how symmetry analysis can determine the number and type of magnetic domains, and provide a practical guideline for selecting optimal altermagnet candidates.

**Altermagnetism- and SOC-induced spin patterns**

MnTe crystallizes in a NiAs-type structure (space group *P6₃/mmc*, #194) and adopts an *A*-type collinear antiferromagnetic ground state, with its Néel vector aligned along $[1\bar{1}00]$ (Section S1 and Figs. S1-S4 in the Supplementary Information details the crystallographic and magnetic characteristics of our MnTe crystals). Its magnetic structure is proposed to feature a "*g*-wave altermagnetic" spin texture with the electron spins of all Bloch states aligning along the Néel vector. This spin texture is accompanied with giant spin splitting (~1 eV) off the high symmetry momenta[8–12]. Governed by the spin group symmetries[1,30–34] $[C_2||C_{6z}t]$ and $[C_2||M_z]$, the spins change signs upon six-fold rotation in the momentum space or going across the $k_z = 0$ plane. For the tangential component of in-plane spin ($S_t$), an additional sign change occurs with respect to $k_x = 0$ due to $[TC_2||C_{2x}]$ (Fig. 1**c**). This leads to a sign change in the spin polarization along three consecutive *B-O-B* paths, as $S_t$ goes from zero at 0° with respect to the vertical direction in Fig. 1**c**, to a "+ –" pattern at 60°, and to a "– +" pattern at 120° (Fig. 1**e**) (See also Sections S4-S5 and Figs. S8-S9 in the Supplementary Information). On the other hand, SOC perturbatively induces band splitting along all high symmetry lines.

In realistic materials where SOC leads to magnetic anisotropy, magnetic domains generally form because multiple symmetry-connected magnetic configurations share a degenerate ground state energy[13]. Starting from the nonmagnetic grey group $P6_3/mmc1'$ of MnTe, the experimentally confirmed Néel vector corresponds to the order-6 magnetic subgroup $Cm'c'm$, indicating that six equivalent symmetry-broken magnetic domains could exist. Among them, three are related by the 3-fold rotation, and the other three are obtained by applying time reversal. Meanwhile, each domain is inversion symmetric (For details, see Section S7 and Fig. S11 in the Supplementary Information). Consequently, in an SARPES experiment, when the spot size of the incident beam is considerably larger than the domain size, the measured altermagnetic spin polarization should be canceled by the domains.

Therefore, although the altermagnetic spin splitting of MnTe persists even in the presence of magnetic domains, any domain-averaged spin polarization in MnTe, if existing, should originate from the surface SOC with inversion symmetry inherently broken. The corresponding spin polarization, depicted in Figs. 1**f-h** (and detailed in Section S8, with Figs. S13-S16), fundamentally deviates from the altermagnetism physics (Figs. 1**c-e**) by the following three aspects: Firstly, the domain-averaged spin texture no longer exhibits the alternating spin reversal associated with the altermagnetic order (Fig. 1**c**). Instead, it shows a Rashba-like texture where the in-plane spin rotates persistently clockwise or counterclockwise along both $\overline{\Gamma}$-$\overline{K}$ and $\overline{\Gamma}$-$\overline{M}$ (Fig. 1**f**, also see Methods for the $k \cdot p$ Hamiltonian of the surface states). Secondly, the spin polarization is significantly stronger near the Fermi level (surface states) and becomes notably weaker at higher binding energies (bulk states) (Fig. 1**g**). Thirdly, the band splitting is no longer confined to specific symmetry planes in the Brillouin zone—it persists even within the Γ*MK* and Γ*KHA* mirror planes. These features, clearly distinguishing from an altermagnetic origin, are all observed by our SARPES measurements, as will be discussed in the following.

### $k_∥$- and $k_z$-independent Rashba-like spin texture near $E_F$

For SARPES experiments, we first measure the spin polarization near the Fermi level, as spins in this energy region are most closely related to the quantum transport behavior. From Fig. 2**b**(i), we notice that four sharp, hole-like bands emerge at binding energies 0-0.5 eV. The momentum locations and dispersions of these bands resemble our slab calculation results in Fig. 1**g** rather than the bulk band structure in Fig. 1**d**, hinting their surface origin. To examine the spin texture of these bands, we selected $hv$ = 108 eV ($k_z$ = 11.5 $\pi/c$) and performed spin-MDC measurements along $\overline{\Gamma}$-$\overline{M_1}$ ($O_1$-$B_1$ in bulk band notations, denoted as Cut 1). The spin component $S_t$ is measured, which points into ("+") or out of ("–") the page in Fig. 2**b**(i). Clear $S_t$ polarization signal is observed in the four bands, with a conclusive "+ – + –" alternating pattern and polarization magnitudes up to 25%.

Since a spin pattern antisymmetric with respect to $\overline{\Gamma}$ along Cut 1 is consistent with both the altermagnetism-induced, "plaid-like" spin texture as well as the SOC-indued, Rashba-like spin configuration, one must measure $S_t$ on other momentum directions to distinguish between the two scenarios. To do this, we first rotated the measurement direction from $\overline{\Gamma}$-$\overline{M_1}$ to $\overline{\Gamma}$-$\overline{M_2}$ [from $O_1$-$B_1$ (Cut 1) to $O_1$-$B_2$ (Cut 2)] while keeping all other experimental parameters unchanged. According to Fig. 1**e**, if altermagnetism were the origin of spin splitting in these bands, the $S_t$ pattern would either becomes zero everywhere (0°) or change to "– + – +" (120°); if the spin polarization is Rashba-like, the pattern will be the same as that in Cut 1. As shown in Fig. 2**b**(ii), the latter is clearly the case. The sign of the $S_t$ polarization does not vanish or reverse compared to Cut 1. This behavior contrasts sharply with the in-plane spin texture predicted within the altermagnet framework (Figs. 1**c-e**), but aligns perfectly with the Rashba-like spin pattern predicted by the SOC-driven model (Figs. 1**f-h**). Due to matrix element effects, the two inner bands appear stronger along Cut 2 than those along Cut 1, while the two outer bands appear weaker along Cut 2, yet the "+ – + –" spin polarization remains undeniable for both cuts.

Next, we change the incident photon energy in the SARPES measurements to examine whether the spin pattern is $k_z$-dependent, as predicted in the altermagnet model of MnTe. These measurements were performed along Cut 3 (Γ-*M*, $h\nu$ = 120 eV, $k_z$ = 12 $\pi/c$) and Cut 4 ($O_2$-$B_3$, $h\nu$ = 129 eV, $k_z$ = 12.5 $\pi/c$). According to spin group symmetry, if the spin splitting arose from the altermagnetic order, mirror symmetry would enforce spin degeneracy at $k_z$ = $n\pi$ [$n$ = 0, 1, 2, …], and polarization of the in-plane spin would be antisymmetric with respect to these planes. However, as shown in Figs. 2**b**(iii)-(iv), the spin pattern of the near-$E_F$ bands remains essentially unchanged across the two sides of $k_z$'s within a Brillouin zone, showing the persistent "+ − + −" configuration.

To further explore the spin texture across different binding energies, we performed spin-MDC measurements at $E_B$ = 0.5 eV for Cuts 1 and 2. As shown in Fig. 3**b**, these spin-MDCs exhibit consistent spin polarization signs ("+ + − −") along different momentum direction, which is nicely reproduced in our slab-model calculation. This consistency suggests that the electronic states at $E_B$ = 0.5 eV likely originate also from the surface states, and thus share similar, Rashba-like characteristics with the bands near the Fermi level.

**Spin texture of the bulk states**

To further examine whether the bulk spin texture of MnTe exhibits characteristics of altermagnetism, we performed SARPES measurements on bands located at higher binding energies, where bulk state contributions are expected to dominate. First, when comparing the spin-MDCs at different binding energies within Cut 2 [Fig. 3**b**(ii)-(iii)], we note that the $S_t$ polarization at $E_B$ = 2 eV is significantly reduced compared to that at $E_B$ = 0.5 eV, and ceased to display a clear polarization pattern. This suggests that the bands at $E_B$ = 2 eV are not directly related to the surface-derived states near the Fermi level. This interpretation is supported by calculations shown in Fig. 1**g**, where states at $E_B$ < 1 eV are shown to be dominated by surface contributions, whereas those at $E_B$ > 1 eV are primarily of bulk origin. Fig. S6 in the Supplementary Information presented additional spin-MDCs measured at $E_B$ = 1.2 eV and $k_z$ = 9.5 $\pi/c$. There we see no clear signature for sign reversal of $S_t$ between two $\bar{\Gamma}$-$\bar{M}$ cuts that are 60° apart azimuthally. Using the same logic as in Fig. 2, we could not identify conclusive $S_t$ polarization that are of altermagnetic origin in these measurements.

In Fig. 4 we elaborate the in-plane spin texture at these high binding energies with another dataset taken at a different facility (see Methods). In particular, the spin-MDCs were extracted by integrating the SARPES signal within a wide binding energy window of 1.0–1.8 eV, covering a large energy region where little contribution from the surface states is expected (Fig. 1**g**). As shown in Fig. 4**b**, both the spin-EDCs and the spin-MDCs hint that the $S_t$ polarization of the bands within these binding energies is for a large part positive for both $k_{\bar{\Gamma}\bar{M}} < 0$ and $k_{\bar{\Gamma}\bar{M}} > 0$, and shows no apparent dependence on $k_z$. Such "all-positive" $S_t$ pattern is seemingly inconsistent with both the altermagnetic and the Rashba-like spin texture. However, due to the low signal-to-noise ratio, no conclusive spin polarization pattern can be deduced at these binding energies. Fig. S7 in the Supplementary Information presented additional SAPRES data obtained along the $\bar{\Gamma}$-$\bar{K}$ direction. There again, no clear signal of $S_t$ polarization can be seen. Therefore, as a corollary, no conclusive signature of spin polarization is currently available in its bulk bands allocated further away from $E_F$.

**Symmetry-related multi-domain**

From the perspective of symmetry, we now discuss the role of antiferromagnetic multi-domain that leads to the compensation of spin polarization. Notably, due to the small domain size[13] (rather than an issue with data quality), neither the SARPES dataset from this work nor that from Ref. 10 points to clear conclusions on whether the MnTe bulk bands exhibit the characteristic *g*-wave altermagnetic spin texture. In contrast, relatively apparent spin texture is observed in $MnTe_2$, candidate of noncollinear spin-split antiferromagnet, in which the domain size is observed to be comparable with the beam spot[35]. The comparison between $MnTe_2$ and MnTe illustrates how symmetry governs magnetic domain formation. In $MnTe_2$, the crystallographic space group, magnetic space group, and spin space group happen to be the same ($Pa\bar{3}$), enforcing the system to possess only two time-reversal-related domains. In contrast, the easy-plane order in MnTe breaks the six-fold screw rotation symmetry and produces six equivalent domains with moments rotated by multiples of 60°, making domain formation energetically more favorable and directly influencing its spectroscopic signatures.

Another interesting example is CrSb, which shares the same lattice as MnTe, but has a different Neel-vector orientation[36–40]. CrSb is an easy-axis magnet (magnet with moments along the principal axis of the point group) that preserves half of the parent crystal symmetries, also allowing only two time-reversal-related domains. Since domain formation in this case requires complete moment reversal, corresponding to a relatively high energy barrier, it is far less favorable than that in MnTe. This explains why altermagnetic spin textures are more convincingly observed in CrSb[40].

In general, the number of degenerate magnetic domains is determined by how the crystal symmetry is broken by the magnetic order. Landau theory predicts that magnetic structures transforming under one-dimensional and real-valued irreducible representations of the parent grey group host only two domains, while higher-dimensional representations allow multiple domains with relative orientations less than 180°. In particular, all 16 abelian crystallographic point groups ($C_n, C_{nh}, S_{2m}, C_{2v}, D_2, D_{2h}, n = 1, 2, 3, 4, 6, m = 1, 2, 3$) support only time-reversal-related domains, whereas the others permit either two or $2n$ ($n > 1$) domains. Such symmetry-based classification provides a practical guideline for selecting candidate altermagnets and other unconventional magnets, where excessive domain formation may suppress physical responses such as spin polarization and spin Hall currents.

**Discussion**

Although our data on the bulk bands tends to support the origin of spin polarization other than altermagnetism, we cannot rule out the possibility that an antiparallel spin pattern resembling that in an altermagnet might be obtained at a different experimental setting (photon energy, light polarization, beam spot size, surface strain, etc), where spin signal from domains with a certain moment direction is favorably captured by SARPES. For spintronic application of altermagnets, a universal question on multi-domain altermagnet candidates then arises: how may one retrieve the altermagnetic spin polarization texture via the control of AFM domains? Here we propose a general strategy for the emergence and observation of altermagnetic spin textures.

Although altermagnetism is formally defined in the absence of SOC—under which MnTe would host only in-plane spin components parallel to the magnetic moments—the inclusion of SOC introduces an additional out-of-plane spin polarization ($S_z$) beyond the altermagnetic spin texture. The magnetic symmetry of MnTe allows for small $M_z$ components in each magnetic moment, implying that the $S_z$ spin texture cannot be canceled by the three domains related by rotations about the $z$-axis. In contrast, the other three time-reversal-related domains manifest opposite spin components. Consequently, an external magnetic field applied along the $z$-axis can lift the energy degeneracy between these domains, stabilizing a single rotationally related domain set—a behavior as observed experimentally[13]. Therefore, for MnTe and other easy-plane altermagnets with symmetry-allowed $M_z$ components and anomalous Hall effect, the $S_z$ spin texture should emerge once a magnetic field selects a rotation-related domain configuration.

We calculated the $S_z$ spin texture for both single-domain and triple-domain bulk states. In the single-domain case, the $S_z$ components are even functions of the $k_x$, $k_y$, and $k_z$ directions, consistent with the magnetic point group $m'm'm$ (see Fig. S9**j-l**). At a binding energy $E_B$ = 0.5 eV, the spin texture reverses sign between $k_x$ and $k_y$ directions. In the triple-domain configuration, the $S_z$ components are invariant under 6-fold rotations, consistent with the $6/mm'm'$ symmetry (see Fig. S12). Our calculation further shows that the $S_z$ spin texture remains nearly unchanged with in-plane momentum direction, which is another key character expected to be experimentally detected.

In summary, we established via SARPES and DFT calculations that a $k_∥$- and $k_z$-independent, Rashba-like spin polarization pattern exists in the electronic states of MnTe that are likely of surface origin, while no experimental evidence is found for a *g*-wave altermagnetic spin texture in the bulk states. The observed spin texture markedly deviates from the theoretical predictions based on altermagnetism, revealing the complex, SOC-driven nature. Our study delivers the first direct spectroscopic identification of an isotropic spin-polarized state in MnTe, shedding light on the indispensable role of SOC. This insight also paves the way for developing next-generation spintronic devices that exploit SOC-driven phenomena beyond conventional magnetic frameworks.

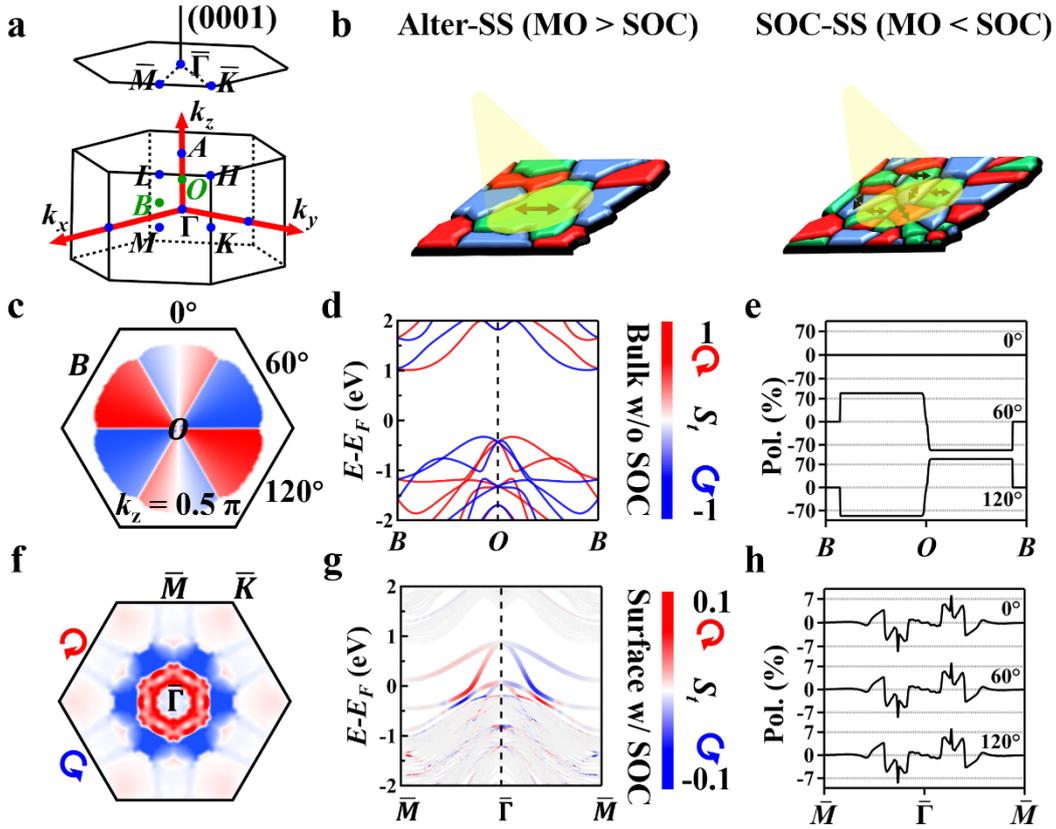

**Fig. 1 | DFT-derived spin patterns for single-domain and multiple-domain configurations in MnTe.** The in-plane spin component perpendicular to the momentum (the tangential component, $S_t$) are shown in red or blue, indicating clockwise or counterclockwise orienting in-plane spins, respectively. **a**, Bulk and surface Brillouin zones (BZ) of MnTe. O and B are mid-points of ΓA and ML, respectively. **b**, Schematic diagrams of the ARPES beam and the magnetic domains for the single-domain (altermagnetic) and multiple-domain (SOC-dominated) cases. MO: magnetic moment magnitude; black double-headed arrows: magnetic moment direction within a domain. **c-e**, The altermagnetic case, obtained via bulk-band calculation. **c**, $S_t$ texture calculated at an experimentally adjusted binding energy (same below) $E_B = 0.5$ eV at $k_z = 0.5$ π/c. **d**, $S_t$-projected bulk bands at $k_z = 0.5$ π/c, along the B-O-B path (parallel to $\bar{M}$-$\bar{\Gamma}$-$\bar{M}$). **e**, Magnitudes of $S_t$ polarization at binding energy $E_B = 0.5$ eV along three inequivalent B-O-B paths (0°, 60° and 120°) marked in **c**. **f-h**, The SOC-dominated case, obtained via a slab-model calculation (See Methods). **f**, $S_t$ texture calculated at $E_B = 0.2$ eV. **g**, Atom- and spin-projected slab bands along $\bar{M}$-$\bar{\Gamma}$-$\bar{M}$ on the top four atomic layers from the surface. The thickness of the bands represents the surface atom projection, and the color indicates the surface spin projection. **h**, DFT-calculated $S_t$ polarization at $E_B = 0.2$ eV along three equivalent $\bar{M}$-$\bar{\Gamma}$-$\bar{M}$ paths (0°, 60° and 120°) marked in **c**. When switching from 0° to 60° and 120° with respect to the vertical direction in **f**, the $S_t$ polarization exhibits an unchanged "+ − + −" pattern.

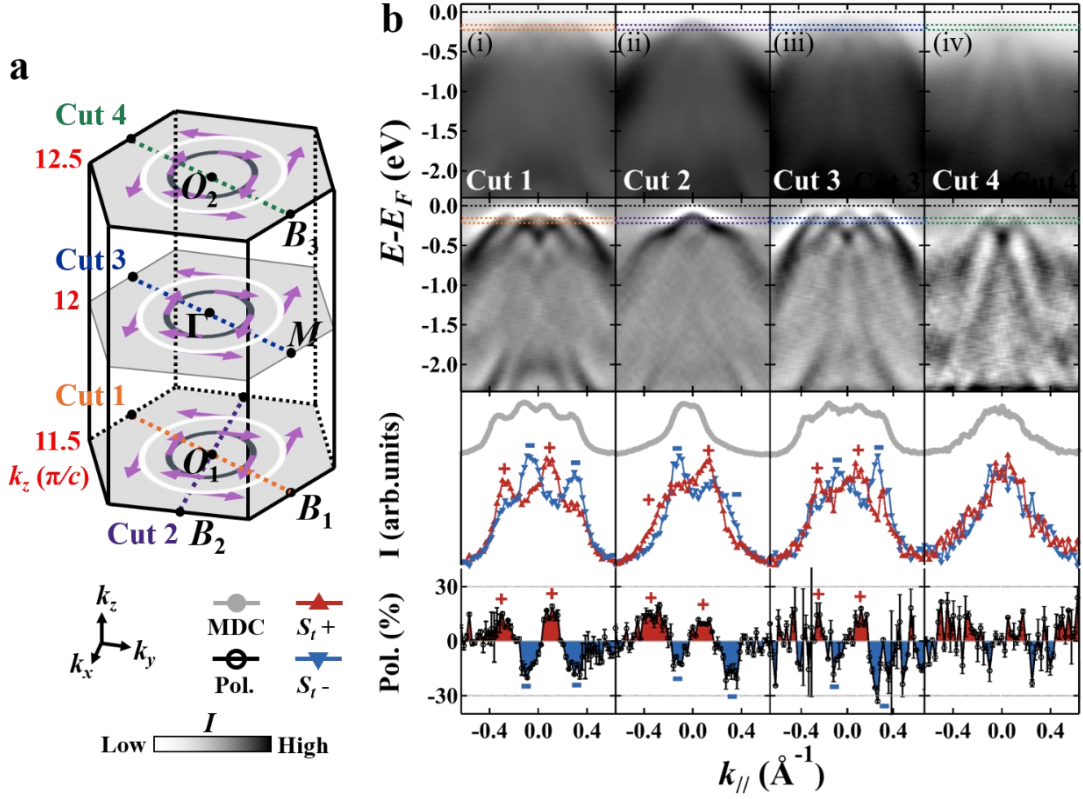

**Fig. 2 | Rashba-like spin texture near the Fermi level. a**, 3D Brillouin zone (BZ) of MnTe with 2D cross-sections showing the $\bar{\Gamma}$-$\bar{M}$-$\bar{K}$ planes at three different $k_z$ values. Colored dash lines define Cuts 1-4 in **b**. Black and white circles centered at $\bar{\Gamma}$ mark the two spin-split bands. Purple arrows represent the experimental in-plane spin directions. **b**, ARPES and SARPES data along Cuts 1-4. First row: spin-integrated ARPES $E$-$k$ maps. Colored rectangles represent the integration area for the spin-MDCs. Second row: corresponding second-derivative maps along the EDCs. Third row: spin-MDC data. Black lines with circles represent the spin-integrated MDC; red lines with up-triangle / blue lines with down-triangle correspond to the $S_t+$ / $S_t-$ intensities. Fourth row: $S_t$ polarization curves. Error bars are defined as the variance within the integrated region; black dashed lines represent the polynomial fitting curves. (i, iii, iv) Cuts 1, 3, 4 acquired at $k_z$ = 11.5, 12, 12.5 $\pi/c$ along $O_1$-$B_1$, $\Gamma$-$M$ and $O_2$-$B_3$. (ii) Cut 2 acquired at $k_z$ = 11.5 $\pi/c$ along $O_1$-$B_2$.

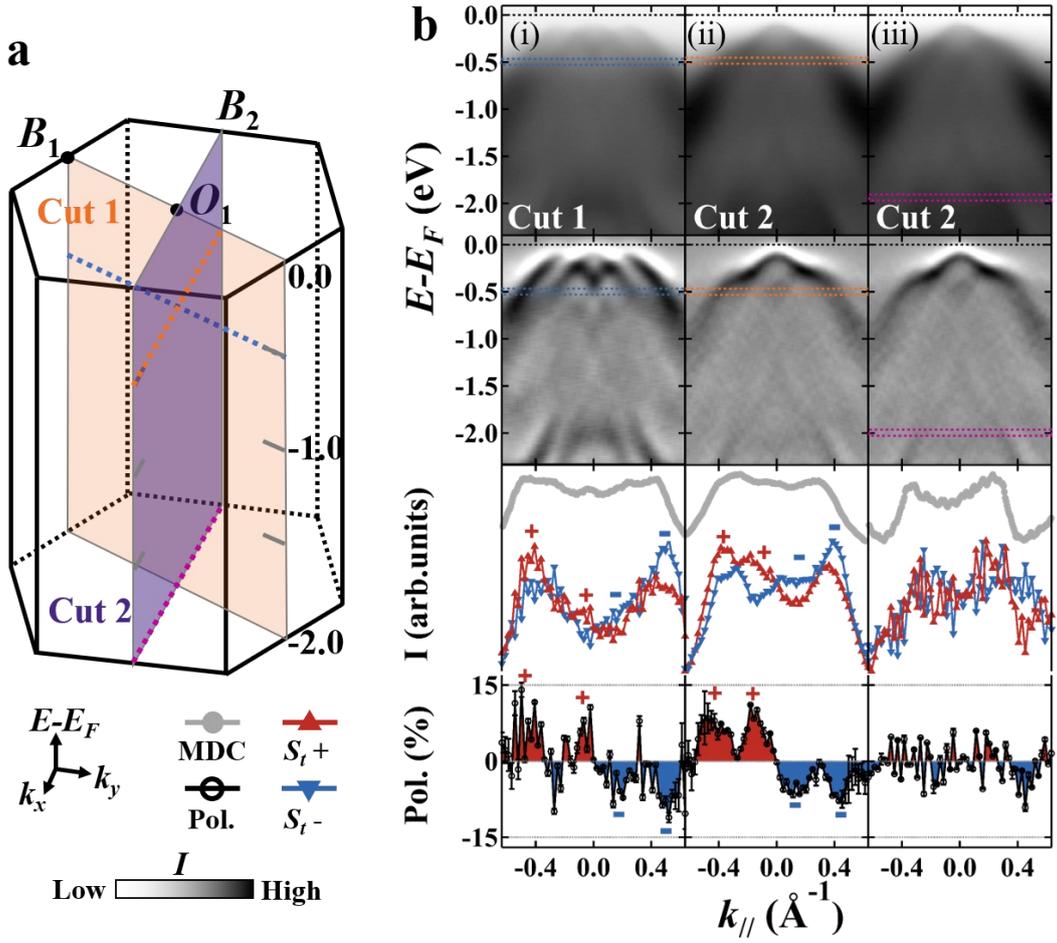

**Fig. 3 | $S_t$ polarization pattern at different binding energies along Cuts 1 and 2 in Fig. 2. a**, A schematic of the 2D BZ with energy as a third dimension. Blue and orange cross-sections correspond to the *k*-positions of Cut 1 and Cut 2; colored dotted lines indicate the spin-MDC measurement positions; **b**, SARPES results at different binding energies along Cuts 1 and 2. Arrangement of panels and definitions of symbols are the same as in Fig. 2.

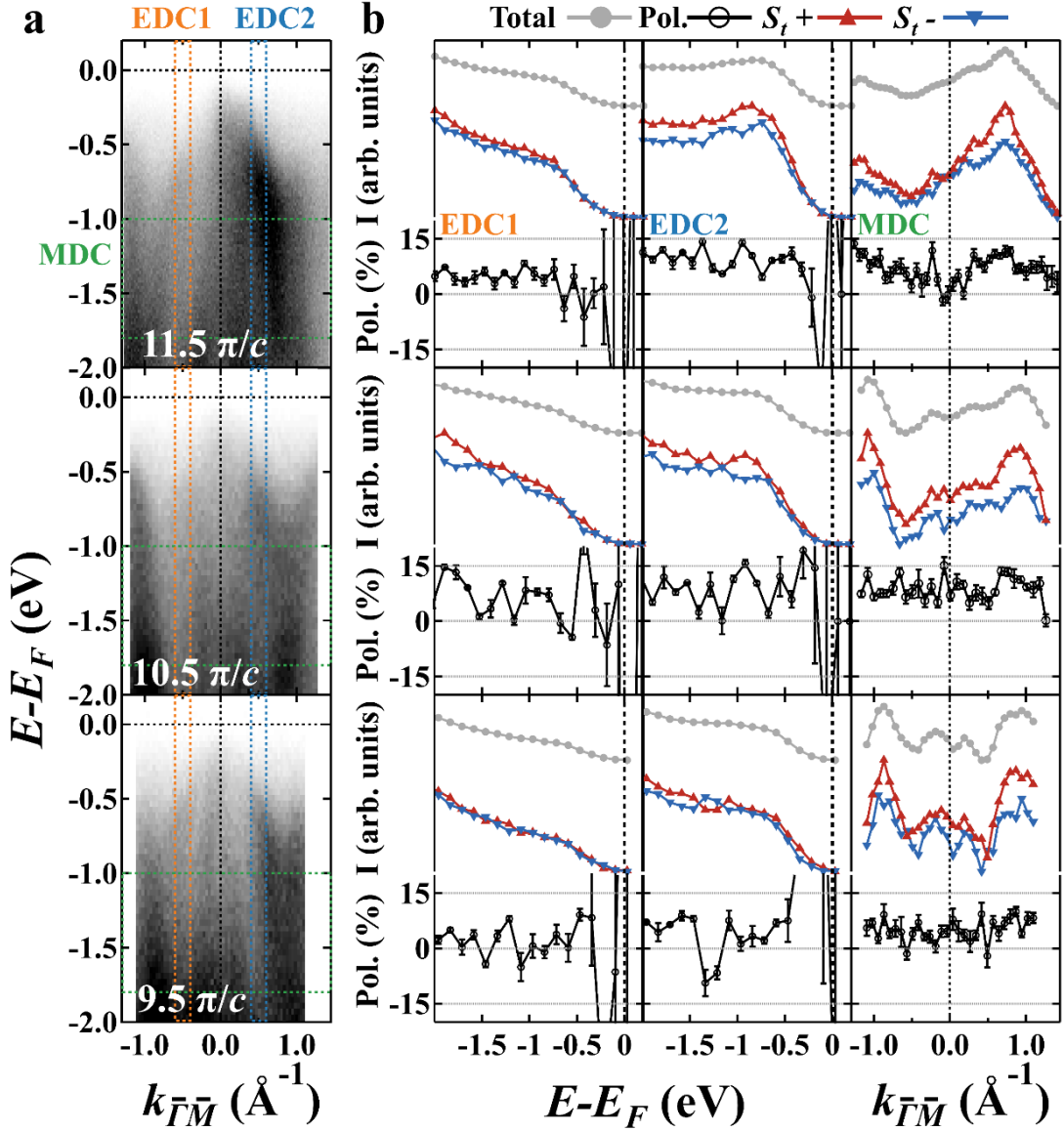

**Fig. 4 | Texture of the in-plane tangential spin $S_t$ in the bulk bands. a,** Spin-integrated ARPES $k$-$E$ maps along $\bar{\Gamma}$-$\bar{M}$, acquired at different $k_z$ positions. **b,** Corresponding $S_t$-polarized EDCs and MDCs. Spin-EDCs are integrated within momentum windows -0.6 < $k_\parallel$ < -0.4 Å$^{-1}$ (orange dashed rectangles in **a**) and 0.4 < $k_\parallel$ < 0.6 Å$^{-1}$ (blue dashed rectangles in **a**). Spin-MDCs are integrated within a binding energy window 1 < $E_B$ < 1.8 eV (green dashed rectangles in **a**), covering a large region where altermagnetic bulk bands are expected to dominate the electronic structure (Fig. 1**e**). Black lines with circles represent the corresponding spin-integrated EDCs and MDCs; red (blue) lines with up (down)-triangle denote the $S_t$+/$S_t$− intensities. Grey lines with circles represent the $S_t$ polarization curves. Error bars are defined as the variance within the respective integration windows. From the data, no signature of spin sign reversal is seen between $k_{\bar{\Gamma}\bar{M}}$ < 0 and $k_{\bar{\Gamma}\bar{M}}$ > 0, as well as between $k_z$ < 0 ($k_z$ = 10.5 $\pi/c$) and $k_z$ > 0 ($k_z$ = 9.5 and 11.5 $\pi/c$).

**Acknowledgements**

Work at SUSTech was supported by the National Key R&D Program of China (Grant No. 2022YFA1403700), the National Natural Science Foundation of China (NSFC) (No. 12534003, No. 12525410), the Guangdong Provincial Key Laboratory for Computational Science and Material Design (No. 2019B030301001), and the Shenzhen Science and Technology Program (Grant No. RCJC20221008092722009). The DFT calculations were performed at Center for Computational Science and Engineering of SUSTech.

**Author Contributions**

C.L. and Q.L. conceived and designed the research project. M.Z. grew and characterized the single crystals. M.Z., X.-R.L., Y.-P.Z., T.-H.S., Y.-J.H., X.-M.M., G.Q., R.K., D.W., R.-H.L., Y.D., X.Z., K.M., K.S., T.O., K.T., Y.H., and C.L. performed the ARPES measurements. P.L., M.-Y.Z., N.Z. and Q.L. performed the theoretical analysis and DFT calculations. M.Z., P.L., Q.L., and C.L. wrote the paper with the help from all authors.

**Data Availability**

The data that supports the findings of this study are available from the corresponding authors on request. Correspondence and requests of ARPES data are addressed to C.L. and those of DFT data are addressed to Q.L.

**Competing Interests**

The authors declare no competing interests.

# Methods

Crystal growth and XRD characterization

Single crystals of MnTe with (000$l$) cleavage plane were grown using the chemical vapor transport (CVT) method. First, the polycrystalline powder was synthesized as a CVT precursor. Starting elements (Mn plates from Aladdin, 99.5% purity; Te ingots from Aladdin, 99.99% purity) were packed in an alumina crucible with a molar ratio of Mn : Te = 1 : 1. The mixture was then sealed in a quartz tube under vacuum. The sealed ampoule was heated for 6 h up to 1000 °C, held for 20 h, then slowly cooled to room temperature over 2 days. Polycrystalline MnTe was then obtained. The powder is subsequently grounded and mixed with $I_2$ (from Aladdin, 99.9% purity) thoroughly in the agate mortar with a molar ratio of MnTe : $I_2$ = 1 : 0.1, and sealed into a silica tube under vacuum. Next, MnTe single crystals were grown using CVT. The sealed ampoule was heated in a two-zone furnace to a low-temperature $T_L$ = 900 °C and a high-temperature $T_H$ = 1000 °C in 24 h and maintained at this condition for two weeks. Millimeter-sized shiny MnTe single crystals were then obtained. The MnTe samples were characterized by XRD at room temperature using a Rigaku SmartLab diffractometer with Cu $K_\alpha$ radiation.

Energy Dispersive X-Ray Spectroscopy Measurement

Energy Dispersive X-ray Spectroscopy (EDX) measurements were performed on a Nova NanoSEM 450 field-emission scanning electron microscope to analyze the chemical composition of the samples. The experiments were carried out under an accelerating voltage of 20 kV and a beam current of 1 nA, providing adequate excitation for reliable detection of characteristic X-ray signals. Prior to measurement, the samples were mounted on conductive carbon adhesive to ensure stable grounding. Both point and area scans were conducted to evaluate the homogeneity and distribution of constituent elements. The acquired spectra were processed using the instrument's dedicated software to confirm the stoichiometry and elemental distribution of the samples.

Magnetic Characterization

Magnetization measurements of the MnTe single crystals were carried out with a superconducting quantum interference device (SQUID)-vibrating sample magnetometer (VSM) system (MPMS3, Quantum design). This system can cool samples down to 1.8 K and generate a variable magnetic field up to ±7 T along both in-plane and out-of-plane directions. FC and ZFC curves were measured by increasing the temperature from 3 to 400 K with in-plane magnetic fields of 500 Oe.

ARPES and SARPES measurements

The data shown in Fig. 2, Fig. 3 and Fig. S6 were acquired at BL09U (Dreamline) of the Shanghai Synchrotron Radiation Facility (SSRF), Shanghai, China, which is equipped with a Scienta-Omicron DA30 electron analyzer and *p*-polarized light. Photon energies used ranged between 50 and 200 eV. The direction of the ARPES exit slit is parallel to $\bar{\Gamma}$-$\bar{M}$. The data shown in Fig. 4 and Fig. S7 were acquired at the URANOS beamline of the SOLARIS National Synchrotron Radiation Centre in Kraków, Poland, which is equipped with a Scienta-Omicron DA30 electron analyzer and *p*-polarized light. Photon energies used ranged between 74 and 129 eV. The slit direction is parallel to $\bar{\Gamma}$-$\bar{K}$, and the $S_t$-resolved ARPES data was acquired by the DA30 deflection mode. Sample temperatures in all measurements were set to be around 30 K. The samples were cleaved *in situ* and measured in a vacuum better than $2 \times 10^{-10}$ Torr. Due to matrix element effects, the signal intensity on the −*k* side is weaker than that on the +*k* side at $k_z = 11.5\ \pi/c$.

The polarization *P* and intensity for $I_\uparrow$ and $I_\downarrow$ in Figs. 2-4 and Figs. S6-S7 are defined as $P = (1/S_{\text{eff}}) \times (I_{\text{mag up}} - I_{\text{mag down}})/(I_{\text{mag up}} + I_{\text{mag down}})$, $I_\uparrow = (1 + P) \times (I_{\text{mag up}} + I_{\text{mag down}})/2$, and $I_\downarrow = (1 - P) \times (I_{\text{mag up}} + I_{\text{mag down}})/2$. The Sherman functions are $S_{\text{eff}} = 0.285$ for data in Figs. 2-3 and Fig. S6, and $S_{\text{eff}} = 0.27$ for data in Fig. 4 and Fig. S7. $I_{\text{mag up}}$ ($I_{\text{mag down}}$) is the photoelectron intensity measured under opposite ferromagnetic target magnetization. The spin-polarized MDCs in Figs. 2-3 and Fig. S6 were obtained by iterating through several loops, in an integration time of about 4 hours. The *P* for the curve obtained in the $n^{\text{th}}$ measurement is denoted as $P_n$, and the error bars are defined as the variance of $P_n$.

Before the SARPES measurements, we performed photon energy (*hν*)-dependent spin-integrated ARPES measurements to accurately identify the high-symmetry planes of MnTe. These measurements were performed at BL09U (Dreamline) of the Shanghai Synchrotron Radiation Facility (SSRF), Shanghai, China, the results of which are presented in Section S2 and Fig. S5. Such experiments are crucial for altermagnet candidates, since spin splitting is expected in an altermagnet only when the momentum deviates from certain high-symmetry directions. These measurements are performed along $\bar{\Gamma}$-$\bar{K}$ using photon energies ranging from 50 to 200 eV. The observed periodic band dispersion agrees well with bulk-band calculations, indicating the precise determination of the $k_z$ value.

DFT calculations

Density functional theory calculations were performed using the projector-augmented-wave method[41] as implemented in VASP[42,43]. The rotational invariant Dudarev's approach[44] to the PBE + *U* method[45] was employed with an effective Hubbard parameter of $U - J = 5$ eV applied to the Mn 3*d* orbitals. For the AFM bulk unit cell, we used a Γ-centered $12 \times 12 \times 8$ *k*-mesh and a 500 eV cutoff. Surface states on the (001) surface were obtained from a slab consisting of 10 layers of the magnetic unit cell stacked along the [001] direction, terminated by Te atoms on both sides by adding an additional Te layer, with a Γ-centered 8 × 8 × 1 mesh and the same cutoff. Band structures of rotation- and time-reversal-related domains were obtained by applying rotational symmetry in reciprocal space and by reversing the Néel vector, respectively.

Momentum-resolved spin distributions were evaluated by Gaussian integration along the energy axis, $\boldsymbol{S}(E_B, \boldsymbol{k}) = \frac{\Sigma_E \boldsymbol{s}(E,\boldsymbol{k}) g(E-E_B)}{\Sigma_E g(E-E_B)}$, where $\boldsymbol{s}(E, \boldsymbol{k})$ denotes the spin polarization of the state with energy $E$ at momentum $\boldsymbol{k}$, and $g(x)$ is the Gaussian function defined as $g(x) = \frac{1}{\sqrt{2\pi\sigma^2}} \exp\left(-\frac{x^2}{2\sigma^2}\right)$. The Gaussian width was set to $\sigma = 0.02$ eV to mimic the energy resolution of ARPES on the order of $10^{-2}$ eV.